\documentclass[apj]{emulateapj}
\usepackage{apjfonts}
\usepackage{amsbsy}

\newcommand{\mpc}{\, {\rm Mpc}}
\newcommand{\hmpc}{{h^{-1}\mpc}}

\begin{document}

\title{Comment on the claimed radial BAO detection by Gazta\~naga et al. }

\author{Jordi Miralda-Escud\'e$^{1,2}$ }
\altaffiltext{1}{Institut de Ci\`encies de l'Espai (IEEC-CSIC),
Campus UAB, Bellaterra, Spain}
\altaffiltext{2}{Instituci\'o Catalana de Recerca i Estudis Avan\c cats,
Barcelona, Spain}

\slugcomment{}
\shorttitle{Comment on Gazta\~naga et al. }
\shortauthors{Miralda-Escud\'e}

\begin{abstract}
Gazta\~naga et al.\ have recently claimed to measure the Baryon Acoustic
Oscillation (BAO) scale in the radial direction from the publicly
available SDSS DR6 data. They focus on the correlation function of
Luminous Red Galaxies (LRG) close to the line-of-sight direction to
find a feature that they identify as the BAO peak, arguing that a
magnification bias effect from gravitational lensing increases the
amplitude of the BAO peak, facilitating its detection. In this Comment,
we clarify that lensing has a negligible impact on the measurement of
the BAO peak, and that the interpretation by Gazta\~naga et al.\ is
incorrect. The feature they identify in the LRG correlation function
near the line-of-sight cannot be explained by any known physical effect
and is in fact consistent with noise.
\end{abstract}


\section{}
\label{sec:intro}

  The Baryon Acoustic Oscillation peak in the galaxy correlation
function was detected for the first time from the Luminous Red Galaxy
(LRG) sample of the Sloan Digital Sky Survey (SDSS) by Eisenstein et
al.\ (2005). This first detection was still at low statistical
significance and was therefore made using the monopole term only, from
the angle-averaged correlation function $\xi(r)$, where
$r^2 = \sigma^2 + \pi^2$, and $\sigma$ and $\pi$ are the comoving
separation across and along the line-of-sight in redshift-space. As the
size of future surveys increases, and the covariance and shot noise
of the measurements of the correlation function near the BAO scale are
reduced, the BAO peak should be fully detectable in redshift-space and
its predicted dependence on the angle as well as $r$ should become
testable.

  Recently, Gazta\~naga, Cabr\'e, \& Hui (2008, hereafter GCH)
have examined the redshift-space LRG correlation function from the Data
Release 6 of SDSS, as a function of both $\pi$ and $\sigma$. In their
Figures 6 and 7, the Kaiser effect (which squashes the contours of the
correlation function owing to peculiar velocities in the linear regime;
see Kaiser 1987) seems clearly present out to scales approaching the BAO
peak, and the BAO peak seems to be present over the whole sphere in
redshift-space. Future analyses should examine the significance at which
the BAO peak can be independently detected in different angular
intervals. Our comments in this note will be restricted to the claim by
GCH of a detected BAO peak close to the line-of-sight.

  In their study, GCH focus their attention to the LRG correlation
function in a very narrow region ($\sim 0.05$ radians) close to the
line-of-sight, to claim the detection of a BAO peak within this region
with an amplitude that is much higher than the theoretical expectation.
They attribute the large amplitude of their claimed BAO peak to a
gravitational lensing magnification effect. They then measure the
central position of the peak, which leads them and Gazta\~naga, Miquel,
\& S\'anchez (2008) to infer various cosmological implications. The
purpose of this note is to clarify that there is no lensing
magnification effect that can appreciably impact the BAO peak
measurement, and that the feature in the correlation function
near the line-of-sight pointed to by GCH as a BAO peak cannot be due
to any known physical effects and is consistent with noise.

  GCH measure the LRG correlation function near the line-of-sight by
averaging over square pixels with side $\Delta\sigma = \Delta \pi = 5
\hmpc$, centered at $\sigma = 3 \hmpc$ and varying $\pi$. They find
a peak of this correlation function at $\pi \simeq 110\, h^{-1}
{\rm Mpc}$ with an amplitude $\Delta \xi \simeq 0.05$ (see their Figs. 8
and 12). The theoretically expected BAO peak amplitude for a galaxy bias
factor $b=2$ (approximately the correct value for LRGs in the SDSS;
e.g., Eisenstein et al.\ 2005) is about 10 times smaller. GCH
nevertheless claim that the peak they find is the BAO peak, attributing
this factor of 10 discrepancy in the peak amplitude to a non-linear
gravitational lensing effect. However, as discussed in Yoo \&
Miralda-Escud\'e (2008, hereafter YM08), the effects of gravitational
lensing on the BAO peak are negligible, causing changes in the position
and amplitude of the peak of $\sim$ one part in $10^4$ only.

  We first clarify the gravitational lensing effect expected in linear
theory. The cross-correlation of galaxy density and magnification bias
by lensing contributes an additive term to the measured correlation
function of LRGs of luminosity greater than $L_t$ at redshift $z$ which,
close to the line-of-sight ($\pi \gg \sigma$), is approximated by
\begin{equation}
 \xi_{gl}(\sigma, \pi) = 3H_0^2 \Omega_m \alpha (1+z)\, bc_{gm}\,
 \pi\, w_p(\sigma) ~,
\end{equation}
where the slope $\alpha = - d\log \bar n_g/ d\log L - 1$,
$\bar n_g (L,z)$ is the number density of galaxies at redshift $z$ with
luminosity above $L$ (the derivative in $\alpha$ is evaluated at the
luminosity threshold $L_t$ for inclusion in the survey), $b$ is the bias
factor, $c_{gm}$ is a galaxy-mass cross-correlation bias, and $w_p$ is
the projected mass correlation function. For the SDSS LRG spectroscopic
sample, the value of the slope is $\alpha\simeq 2$, as inferred for
example using the Brown et al.\ (2007) luminosity function to evaluate
$\alpha$ at the threshold luminosity $L_t = 3L_*$ (which yields a LRG
density $\bar n_g(L_t) \simeq 10^{-4} h^3 \mpc^{-3}$). The
cross-correlation bias $c_{gm}$ should be close to unity on the scales
that are relevant here, which are larger than the size of virialized
halos, unless large-scale galaxy fluctuations are caused by other
factors in addition to mass fluctuations. Using $b=2$ and $c_{gm}=1$,
one finds that $\xi_{gl}$ at the mean redshift of the SDSS sample,
averaged within a transverse distance $\sigma < 5.5 h^{-1} \mpc$, has a
value $\xi_{gl} \simeq 10^{-3}$ at the BAO scale, $\pi = r_{BAO}$, and
is a slowly varying function of $\pi$ (see Figs. 2 and 3 in YM08).

  GCH find a value for $\xi_{gl}$ larger by a factor of several because
they use $b=5.8$ and $\alpha=2.75$ instead (in the notation of GCH,
$s=(\alpha + 1)/2.5 = 1.5$; they use $b=2$ for the galaxy correlation
but $b=5.8$ for $\xi_{gl}$). The observed LRG correlation
function is consistent with a roughly constant bias down to the scales
that are relevant here, and one should use a consistent value of the
bias factor for computing the galaxy correlation function and the
galaxy-magnification cross-correlation. The value of $\alpha=2.75$ used
by GCH is obtained from the overall slope of the galaxy counts at all
redshifts at a fixed apparent magnitude, but the slope of the luminosity
function at a fixed redshift should be used instead because the
correlation function is measured at a specific redshift from a
spectroscopic sample. Even the overestimated lensing effect of GCH is
still much smaller than the amplitude of their claimed BAO peak.
Moreover, the lensing effect adds a slowly varying function to the
galaxy correlation, which does not appreciably change the amplitude or
position of the BAO peak. We note in particular that the galaxy
correlation function happens to be close to zero near the
line-of-sight and at the maximum of the BAO peak, and so the lensing
contribution may appear to be large as a fractional change of
$\xi(r_{BAO})$; this fractional change is however irrelevant for the
purpose of measuring the BAO peak. It is shown in YM08 that the actual
fractional amount by which the position and amplitude of the BAO peak
are modified by lensing at $z \sim 0.3$ is $\sim 10^{-4}$ within 15
degrees of the line-of-sight, increasing rather slowly as the distance
to the line-of-sight is reduced.

  GCH argue for the presence of non-linear effects arising from the
coupling of galaxy peculiar velocities and lensing magnification.
Actually, no such effect may possibly introduce new terms in the
correlation function that are larger than the linear effect calculated
from equation (1). The lensing contribution $\xi_{gl}$ arises from the
correlation of the magnification bias on one line-of-sight with the
galaxy density on the other. This is calculated using the
non-linear mass correlation $\xi_{mm}$, so the only non-linear effects
that are not included are the peculiar velocities on the line-of-sight
where the galaxy density is evaluated. Non-linear peculiar velocities
can only redistribute the galaxies along the line-of-sight over a scale
that is much smaller than the BAO scale $r_{BAO}$, but cannot change
the total integrated number of galaxies. Any effects in the galaxy
density caused by this redistribution along the line-of-sight become
washed out by the integration that is involved in computing the lensing
magnification.

  Furthermore, the correctness of the theoretical calculation of the
lensing effects on the galaxy correlation function using equation (1) is
observationally tested by measurements of the average lensing shear
around foreground galaxies (Sheldon et al.\ 2004), as discussed in YM08.
This confirms that lensing can introduce only tiny corrections for the
measurement of the BAO peak.

  In agreement with the GCH analysis, the peak in the LRG correlation
function along the line-of-sight near the BAO scale found in the SDSS
DR6 is consistent with a noise spike with a probability of $\sim 2\%$ of
occurring randomly. For the purpose of estimating this probability, it
is sufficient to roughly evaluate the shot-noise contribution to the
correlation function error: the number
density of LRGs used in SDSS is $n\sim 10^{-4} h^3 \mpc^{-3}$, and the
volume of the pixels used by GCH to measure the correlation function is
$\sim 500 (\hmpc)^3$, so each LRG has an average number of pairs in each
pixel of $\sim 0.05$. The total number of LRGs used is
$\sim 7\times 10^4$, so the average number of pairs contributing to a
pixel is $\sim 3000$. This yields a relative error due to shot-noise of
$(3000)^{-1/2} \simeq$ 2\% for the correlation function in one pixel,
assuming a homogeneous selection function over the survey volume. The
values of the correlation in different pixels are of course correlated.
The shot noise in this case is dominant, but may be increased by the
tendency of LRGs to occur in massive clusters (implying that the
correlation function errors may be affected by the presence of a few
pairs of massive clusters in the survey with a separation that happens
to be close to the BAO scale along the line-of-sight), and by an
inhomogeneous galaxy density (due to sampling selection) in the survey.
This rough estimate is consistent with the errors calculated in GCH (see
their Figs. 8 and 11). In fact, GCH admit that the probability for the
peak occurring randomly in their full sample is $3\%$, as obtained from
simulations that use the full selection function.

  It is important to note that
this probability should be considered to be {\it a posteriori}, because
it is only after having noticed the presence of an unexpectedly high
peak in the data for the correlation function near the line-of-sight
that one wonders how to explain this peak. An {\it a posteriori}
probability of 3\% should not be considered statistically significant,
especially taking into account that there are a number of parameters one
may play with, such as the pixel size, the region near the line-of-sight
selected for retrieving the correlation function, and the LRG sample
redshift interval.

  The way GCH obtain a value for the BAO scale $r_{BAO}$ is by using
a smoothed version of the data for the correlation function near the
line-of-sight as a model to fit the data itself. Their small error
for $r_{BAO}$ arises from the fact that they consider a narrow noise
spike to be real, and they use the same narrow spike as their
theoretical model. The apparently high-precision cosmological
constraints obtained by Gazta\~naga, Miquel, \& S\'anchez (2008) are a
consequence of using the small error bar for $r_{BAO}$ obtained with
this invalid method.

\acknowledgements

  J.~M. is supported by the Spanish grants
AYA2006-06341, AYA2006-15623-C02-01, and MEC-CSD2007-00060.


\end{document}